\documentclass[conference]{IEEEtran}
\IEEEoverridecommandlockouts

\usepackage{cite}
\usepackage{amsmath,amssymb,amsfonts}
\usepackage{algorithmic}
\usepackage{graphicx}
\usepackage{textcomp}
\usepackage{xcolor}
\def\BibTeX{{\rm B\kern-.05em{\sc i\kern-.025em b}\kern-.08em
    T\kern-.1667em\lower.7ex\hbox{E}\kern-.125emX}}

\newcommand{\bestMNISTFullAccuracy}{97.2±0.1\%}
\newcommand{\noiselessBestMNISTHundredAccuracy}{90.6±1.7\%}
\newcommand{\noisyBestMNISTHundredAccuracy}{92.0±0.3\%}
\newcommand{\bestFashionMNISTFullAccuracy}{88.0±0.1\%}

\newcommand{\bestBPTTMNISTFullAccuracy}{96.8±0.1\%}
\newcommand{\bestBPTTMNISTHundredAccuracy}{90.2±0.4\%}

\newcommand{\FourBitPhaseQuantisationAccuracy}{89.8±1.5\%}
\newcommand{\TwoBitPhaseQuantisationAccuracy}{57.2±2.5\%}

\newcommand{\bestDeclineNoiseAccuracy}{90.8±0.9\%}
\newcommand{\worstDeclineNoiseAccuracy}{41.0±6.2\%}

\newcommand{\TenBitParameterQuantisationMNISTHundredAccuracy}{89.4±1.5\%}
\newcommand{\EightBitParameterQuantisationMNISTHundredAccuracy}{80.4±2.4\%}

\begin{document}

\title{How to Train an Oscillator Ising Machine using Equilibrium Propagation}

\author{\IEEEauthorblockN{Alex Gower\thanks{This work was supported by UKRI/EPSRC CASE Award under Grant 220191 in partnership with Nokia UK Limited.}}
apg59@cam.ac.uk \\
\IEEEauthorblockA{\textit{Nokia Bell Labs \&} \\
\textit{University of Cambridge}}
}

\maketitle

\begin{abstract}
We show that Oscillator Ising Machines (OIMs) are prime candidates for use as neuromorphic machine learning processors with Equilibrium Propagation (EP) based on-chip learning. The inherent energy gradient descent dynamics of OIMs, combined with their standard CMOS implementation using existing fabrication processes, provide a natural substrate for EP learning. Our simulations confirm that OIMs satisfy the gradient-descending update property necessary for a scalable Equilibrium Propagation implementation and achieve $\sim$\bestMNISTFullAccuracy{} test accuracy on MNIST and $\sim$\bestFashionMNISTFullAccuracy{} on Fashion-MNIST without requiring any significant hardware modifications. Importantly, OIMs maintain robust performance under realistic hardware constraints, including 10-bit parameter quantization, 4-bit phase measurement precision, and moderate phase noise that can potentially be beneficial with parameter optimization. These results establish OIMs as a promising platform for fast and energy-efficient neuromorphic computing, potentially enabling energy-based learning algorithms that have been previously constrained by computational limitations.
\end{abstract}


\section{Introduction}

Equilibrium Propagation (EP) offers a promising alternative to backpropagation for neuromorphic computing through local learning rules that eliminate separate backward passes. By nudging system steady states toward desired outputs, EP is ideal for hardware that naturally performs energy gradient descent toward local minima.

Oscillator Ising Machines (OIMs) are well-suited candidates for EP implementation due to their inherent energy gradient descent dynamics~\cite{basharNoteAnalyzingStability2023}. Unlike discrete Ising solvers~\cite{laydevantTrainingIsingMachine2024}, OIMs feature continuous phase dynamics that enable the precise nudging required by EP, and their coupling strengths and bias fields serve as natural trainable parameters. Particularly promising are CMOS-based OIM implementations using voltage-controlled oscillators (VCOs), which enable precise frequency calibration and eliminate the frequency variability that affects other oscillator systems~\cite{rageauTrainingSynchronizingOscillator2025}.

The synergy between EP and OIMs presents a promising opportunity for fast and energy-efficient neuromorphic computing. Current machine learning hardware faces significant energy consumption challenges, with GPU-based training requiring substantial power~\cite{garcia-martinEstimationEnergyConsumption2019}. By contrast, OIM-based solutions could potentially achieve significant improvements in energy efficiency~\cite{csabaCoupledOscillatorsComputing2020b} and training speed through GHz-frequency energy descent dynamics. This acceleration could revitalize energy-based learning approaches~\cite{assranSelfSupervisedLearningImages2023} that are currently hindered by iterative sampling methods and long relaxation times.

Previous attempts to implement EP on oscillator systems had limited success due to initialization difficulties~\cite{wangTrainingCoupledPhase2024} or frequency synchronization challenges~\cite{rageauTrainingSynchronizingOscillator2025}. Our work presents the first comprehensive framework for implementing EP on existing OIM hardware without major modifications, positioning these devices—originally built for combinatorial optimization—as powerful neural network processors.

\section{Oscillator Ising Machine Dynamics}

An Oscillator Ising Machine (OIM) consists of a network of $n$ coupled self-sustaining nonlinear oscillators with non-varying amplitudes. Over a sufficiently coarse-grained timescale~\cite{main-OIMs-paper}, each oscillator can be parameterized by a phase $\phi_i \in [0,2\pi]$ whose dynamics, for equal frequencies\footnote{Through voltage-controlled oscillator (VCO) calibration~\cite{main-OIMs-paper}, frequency variations can be effectively eliminated in practice.} $\omega_i = \bar \omega$, follow:

\begin{equation}\label{eq:oim-dynamics}
\frac{d\phi_i}{dt'} = -\sum_{j=1, \ j\neq i}^n J_{ij} \sin(\phi_i - \phi_j ) -h_i \sin(\phi_i) - S_i \sin(2 \phi_i),
\end{equation}

where $J_{ij}$ represents coupling strength between oscillators, $h_i$ is the bias field favoring phase alignment with 0 or $\pi$ (defined with respect to a reference oscillator), $S_i$ is the synchronization field encouraging binary phase states, and $t'$ is a dimensionless timescale such that real time $t = t'/\bar\omega$ scales inversely with oscillator frequencies.

Importantly, these dynamics can be reformulated as energy gradient descent dynamics~\cite{basharNoteAnalyzingStability2023}:

\begin{equation}
   \frac{d \phi_i}{dt'} = - \frac{\partial V}{\partial \phi_i},
\end{equation}

for an energy function\footnote{Using symmetric $J_{ij}$ convention with factor $\frac{1}{2}$ to correct for double counting.}:

\small
\begin{equation}
   V = -\frac{1}{2}\sum_{i,j \ i \neq j}^n J_{ij} \cos(\phi_i-\phi_j) - \sum_i^n h_i \cos(\phi_i) - \sum_i^n \frac{S_i}{2} \cos(2\phi_i). 
   \label{eq:oim-energy-V}
\end{equation}
\normalsize

This energy gradient descent property, rare among physical systems, makes OIMs particularly promising for use as neuromorphic machine learning processors.

\section{Equilibrium Propagation}\label{sec:equilibrium-propagation}

Equilibrium Propagation (EP)~\cite{scellierEquilibriumPropagationBridging2017,laborieuxScalingEquilibriumPropagation2020} is a learning algorithm ideal for neuromorphic hardware, as it employs local update rules without the separate backward passes required by conventional backpropagation. It is used to train convergent recurrent neural networks (RNNs) i.e. networks whose dynamics are characterized by the convergence to a stationary point. 

For systems with energy gradient descent dynamics, EP utilizes a `total energy' $F(x,s, \{\theta\})$, which depends on inputs $x$, trainable parameters $\{\theta\}$, and a dynamical state $s$ that includes both hidden and output variables ($s=(h,y)$). The system follows gradient descent dynamics: $\frac{ds}{dt} = - \frac{\partial F}{\partial s}$ (for fixed $x$ and $\{ \theta \}$).

This total energy decomposes as $F = E + \beta \ell$, where $E(x,s,\{\theta\})$ is the free system's energy function, $\ell(y,\hat y)$ is a loss function comparing outputs $y$ to targets $\hat y$, and $\beta$ is the nudging factor that biases dynamics toward configurations that minimize loss.

For each training example $x$, EP proceeds in three phases:
\begin{enumerate}
  \item \textbf{Free phase:} Initialize at a reference state $s_0$ with $\beta=0$ and evolve to stationary point $s_*$.
  \item \textbf{Positive nudged phase:} From $s_*$, evolve with $\beta > 0$ to reach stationary point $s_*^{\beta}$.
  \item \textbf{Negative nudged phase:} From $s_*$ again, evolve with $\beta < 0$ to reach stationary point $s_*^{-\beta}$.
\end{enumerate}

The updates to the trainable parameters $\{ \theta \}$ are then given by~\cite{laborieuxScalingEquilibriumPropagation2020} $\Delta \theta = \eta \hat{\nabla}^{\rm EP}(\beta)$ for learning rate $\eta$ and where:

\begin{equation}\label{eq:ep-parameter-update}
    \hat{\nabla}^{\rm EP}(\beta) = -\frac{1}{2\beta} \left( \frac{\partial F}{\partial \theta}(x, s^{\beta}_*, \{\theta\}) - \frac{\partial F}{\partial \theta}(x, s_*^{-\beta}, \{\theta\}) \right).
\end{equation}

This update effectively steers the free stationary point toward configurations with lower loss by decreasing energy at $s^\beta_*$ and increasing it at $s^{-\beta}_*$. In the limit of infinitesimal $\beta$, this update equals gradient descent on the loss~\cite{laborieuxScalingEquilibriumPropagation2020}:
\begin{equation}
   \lim_{\beta \to 0} \hat{\nabla}^{\rm EP}(\beta) = - \frac{\partial \ell}{\partial \theta}(y_*, \hat y).
\end{equation}

Moreover, during nudged phase dynamics before convergence, the instantaneous updates match those from Backpropagation Through Time (BPTT)~\cite{werbosBackpropagationTimeWhat1990}:

\small
\begin{equation}
 \lim_{\beta \rightarrow 0} -\frac{1}{2\beta} \left( \frac{\partial F}{\partial \theta}(x, s^{\beta}_t, \{\theta\}) - \frac{\partial F}{\partial \theta}(x, s_t^{-\beta}, \{\theta\}) \right) = \hat{\nabla}^{\rm BPTT}(t).  
\end{equation}
\normalsize

This additional correspondence reveals that EP computes correct gradients not only at convergence but also during the transient dynamics. Section~\ref{sec:results} demonstrates this correspondence empirically for OIMs, validating their capacity for effective learning despite their distinct periodic coupling structure, and aiding in tuning which finite $\beta$ values to use in practice.

\section{Implementation on Oscillator Ising Machines}\label{sec:implementation-on-oims}

Here we implement a neural network with dense layers on an OIM with $n_{x}$ input neurons (fixed during dynamics), $n_h$ hidden neurons\footnote{For simplicity, we demonstrate the case with one hidden layer, though the implementation extends to arbitrary numbers of hidden layers.}, and $n_y$ output neurons. Each non-input neuron corresponds to an oscillator, with the dynamical system state represented as $\phi = (\phi^{(h)},\phi^{(y)}) = (\{ \phi_i^{(h)} \}_{i=1}^{n_h}, \{ \phi_i^{(y)}\}_{i=1}^{n_y})$. The trainable parameters consist of interlayer weights and biases: $\{ \theta \} = ( \{ w_{ij}^{(x,h)} \}, \{ w_{ij}^{(h,y)} \}, \{ b_i^{(h)} \}, \{ b_i^{(y)} \})$.

We implement neural network components through specific energy terms in the OIM: biases via $-b_i \cos(\phi_i)$ terms, and hidden-to-output weights via $-w^{(h,y)}_{ij} \cos(\phi_i^{(h)} - \phi_j^{(y)})$ terms. Using the mapping $s_i = \cos(\phi_i)$ to transform phases into values in $[-1,1]$, we implement input-to-hidden weights through $-w^{(x,h)}_{ij} x_i \cos(\phi^{(h)}_j)$ terms.

For an MSE loss function $\ell(y,\hat y) = \frac{1}{2} \sum_i(y_i-\hat y_i)^2$ with $y_i=\cos(\phi^{(y)}_i)$, each term expands as:
\begin{align}
    \frac{1}{2}\big( \cos(\phi^{(y)}_i) - \hat y_i \big)^2 &= \frac{1}{4}\cos(2\phi_i^{(y)}) - \hat y_i\cos(\phi^{(y)}_i) + \text{const.} \label{eq:mse_end}
\end{align}
where we used the identity $\cos^2(x) = \frac{1}{2}[1 + \cos(2x)]$ and collected constants that don't affect dynamics.

All energy terms detailed above align with the OIM's energy function $V$ in \eqref{eq:oim-energy-V}, allowing us to map the total energy $F = E + \beta \ell$ onto the OIM and perform energy gradient descent $\frac{d \phi_i}{dt} = - \frac{\partial F}{\partial \phi_i}$ by setting:
\begin{align}
    h_i^{(h)} &= b_i^{(h)} + \sum_{j=1}^{n_x} w_{ji}^{(x,h)} x_j \label{eq:h_hidden} \\
    J_{ij}^{(h,y)} &= w_{ij}^{(h,y)} \label{eq:J_hy} \\
    h_i^{(y)} &= b_i^{(y)} + \beta \hat y_i \label{eq:h_output} \\
    S_i^{{(y)}} &= -\frac{\beta}{2} 
\end{align}

Note that input-to-hidden layer weights are implemented through contributions to the \textit{bias} fields $h_i^{(h)}$ of the hidden neurons, eliminating the need for additional oscillators to represent input neurons. The synchronization fields $S_i^{(y)}$ are only required for output neurons in order to implement the MSE loss correctly—it is a fortunate coincidence that these fields are already intrinsic to physical OIM implementations.

With this implementation, the operation of the OIM as a machine learning processor follows the Equilibrium Propagation framework described in Section \ref{sec:equilibrium-propagation}. We initialize the weights using a He initialization scheme~\cite{heDelvingDeepRectifiers2015} and set all biases to zero. Training proceeds over multiple epochs, with each epoch processing mini-batches of training data.

For each training example $x$, we:
\begin{enumerate}
    \item Configure the OIM parameters as specified in \eqref{eq:h_hidden}--\eqref{eq:h_output}
    \item Initialize all oscillators to the reference state\footnote{This corresponds to setting all neuron activations $\cos(\frac{\pi}{2})=0$. We found that any random initialization of phases also works well, provided the same random reference state is used consistently across all dynamics.} $\phi_0 = \{\frac{\pi}{2}\}$
    \item Run the free phase ($\beta=0$) dynamics until convergence to $\phi_*$
    \item Run the positive and negative nudged phase dynamics (initialized at $\phi_*$) until convergence to $\phi_*^{+\beta}$ and $\phi_*^{-\beta}$
\end{enumerate}

Using the Equilibrium Propagation update rule \eqref{eq:ep-parameter-update}, we derive the following parameter updates:

\begin{align}
    \Delta b^{(h)}_{i} &\propto -\frac{1}{2\beta} \Big[\cos(\phi_i^{(h), -\beta}) - \cos(\phi_i^{(h), +\beta})\Big] \label{eq:update_bh} \\
    \Delta b^{(y)}_{i} &\propto -\frac{1}{2\beta} \Big[\cos(\phi_i^{(y), -\beta}) - \cos(\phi_i^{(y), +\beta})\Big] \label{eq:update_by} \\
    \Delta w_{ij}^{(x,h)} &\propto -\frac{1}{2\beta} \Big[x_i\cos(\phi^{(h), -\beta}_j) - x_i\cos(\phi^{(h), +\beta}_j)\Big] \label{eq:update_wxh} \\
    \Delta w^{(h,y)}_{ij} &\propto -\frac{1}{2\beta} \Big[\cos(\phi^{(h), -\beta}_i - \phi^{(y), -\beta}_j) - \nonumber\\
    &\phantom{\propto -\frac{1}{2\beta} \Big[}\cos(\phi^{(h), +\beta}_i - \phi^{(y), +\beta}_j)\Big]
\end{align}

Notably, these updates are entirely local—each parameter update depends only on phases of directly connected oscillators, measurable on-chip at the relevant synapse or neuron. This locality is a key advantage for hardware implementation, eliminating the need for global backpropagation circuitry.

Parameter updates are averaged over each mini-batch before being applied with a learning rate $\eta$. The implementation remains compatible with standard machine learning optimization techniques, including learning rate scheduling, weight decay, and momentum.

\section{Results}\label{sec:results}

\begin{table}[t]

\caption{Hyperparameters for MNIST, Fashion-MNIST (FMNIST) and MNIST/100 experiments}
\centering
\small
\begin{tabular}{lcc}
\hline
\textbf{Parameter} & \textbf{MNIST \&} & \textbf{MNIST/100} \\
& \textbf{FMNIST} & \\
\hline
Free phase integration steps ($T$) & 4000 & 3500 \\
Nudged phase integration steps ($K$) & 400 & 350 \\
Nudging factor ($\beta$) & 0.1 & 0.05 \\
Time step size ($\epsilon$) & 0.45 & 0.5 \\
\hline
\multicolumn{3}{l}{\textbf{Learning rates:}} \\
Hidden layer weights ($\eta_{w}^{(h)}$) & 0.01 & 0.01 \\
Output layer weights ($\eta_{w}^{(y)}$) & 0.001 & 0.001 \\
Hidden layer biases ($\eta_{b}^{(h)}$) & 0.001 & 0.001 \\
Output layer biases ($\eta_{b}^{(y)}$) & 0.001 & 0.001 \\
\hline
Batch size & 128 & 20 \\
Epochs & 50 & 50 \\
\hline
\end{tabular}
\label{tab:hyperparameters}
\end{table}

We validate our Equilibrium Propagation implementation on OIMs using MNIST~\cite{lecunGradientbasedLearningApplied1998} and Fashion-MNIST~\cite{xiaoFashionMNISTNovelImage2017} classification tasks. MNIST consists of 60,000 training and 10,000 test samples of 28×28 grayscale handwritten digits (pixel values $x_i \in [0,1]$), while Fashion-MNIST contains the same structure but with clothing items instead of digits. We also employ MNIST/100, a balanced 1,000/100 train/test subset, enabling direct comparison with previous Ising machine implementations such as D-Wave quantum annealers~\cite{laydevantTrainingIsingMachine2024}.

We implemented the training procedure using PyTorch with Euler integration of the OIM dynamics in \eqref{eq:oim-dynamics}. Table \ref{tab:hyperparameters} details the hyperparameters used.

\begin{figure}[t]
 
  \centering
  
  \includegraphics[width=\columnwidth]{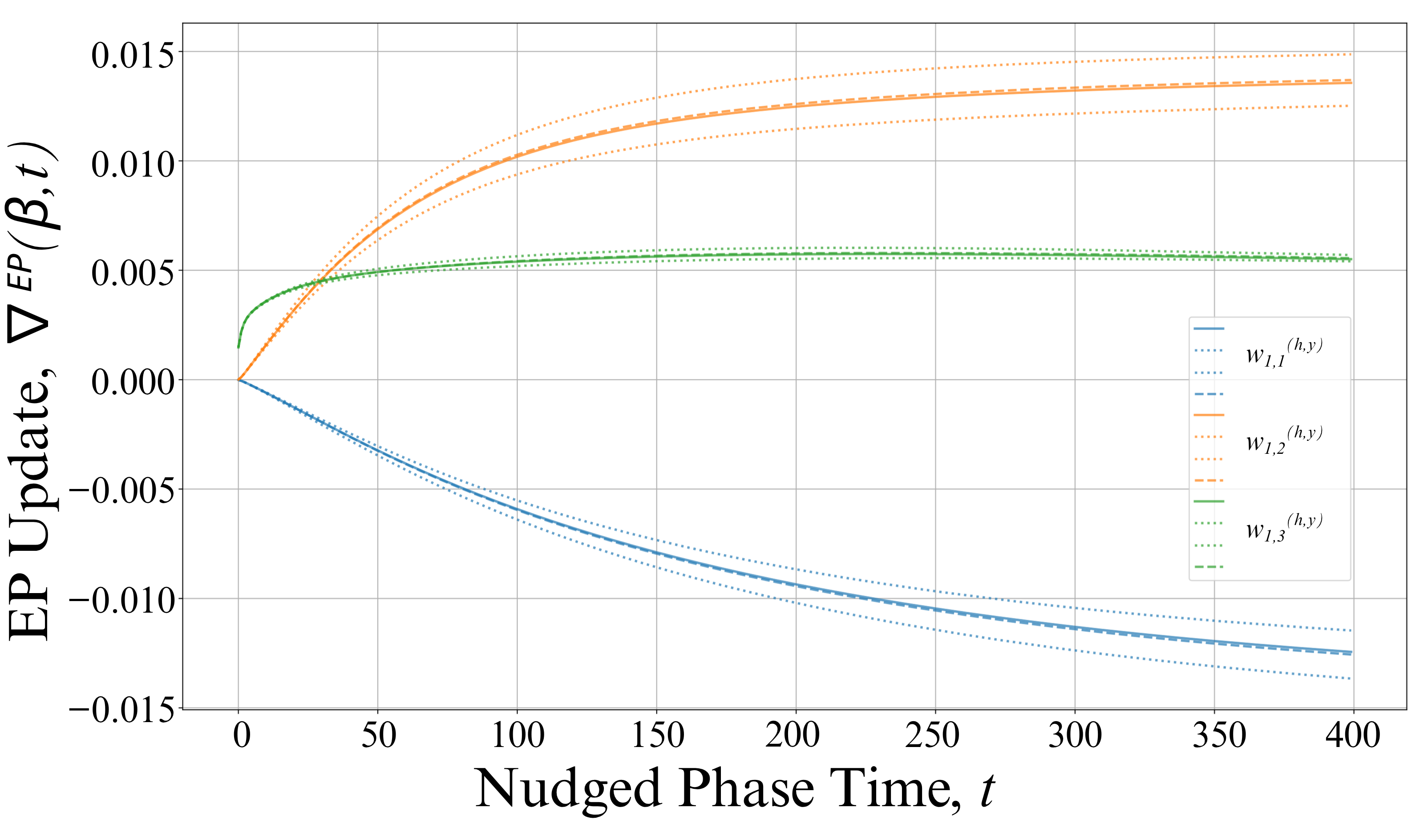}
  
  \vspace{-3mm}
  
  \caption{EP-BPTT correspondence showing symmetric Equilibrium Propagation updates (dashed lines) matching Backpropagation Through Time (solid lines), confirming the gradient-descending update property. Dotted lines show single-phase EP updates using either only positive or only negative nudging~\cite{laborieuxScalingEquilibriumPropagation2020}, demonstrating the advantage of the symmetric approach.}
  
  \label{fig:ep-bptt-matching}
  \vspace{-3mm}

\end{figure}

A critical theoretical validation for EP implementations is the correspondence between EP parameter updates and those from Backpropagation Through Time (BPTT). Fig.~\ref{fig:ep-bptt-matching} demonstrates this match for our OIM implementation. To achieve this correspondence, we tuned hyperparameters by increasing free phase duration $\epsilon T$ and decreasing nudging factor $\beta$ until EP updates matched BPTT updates while maintaining stable dynamics, and ensured the nudged phase duration $\epsilon K$ allowed updates to saturate. This confirms that OIMs satisfy the gradient-descending update property necessary for scaling to complex problems, despite their distinctive energy landscape with trigonometric coupling ($\cos(\phi_i-\phi_j)$) rather than the quadratic coupling ($s_i s_j$) typically used in conventional EP implementations.

\begin{figure}[t]
 
  \centering
  
  \includegraphics[width=\columnwidth]{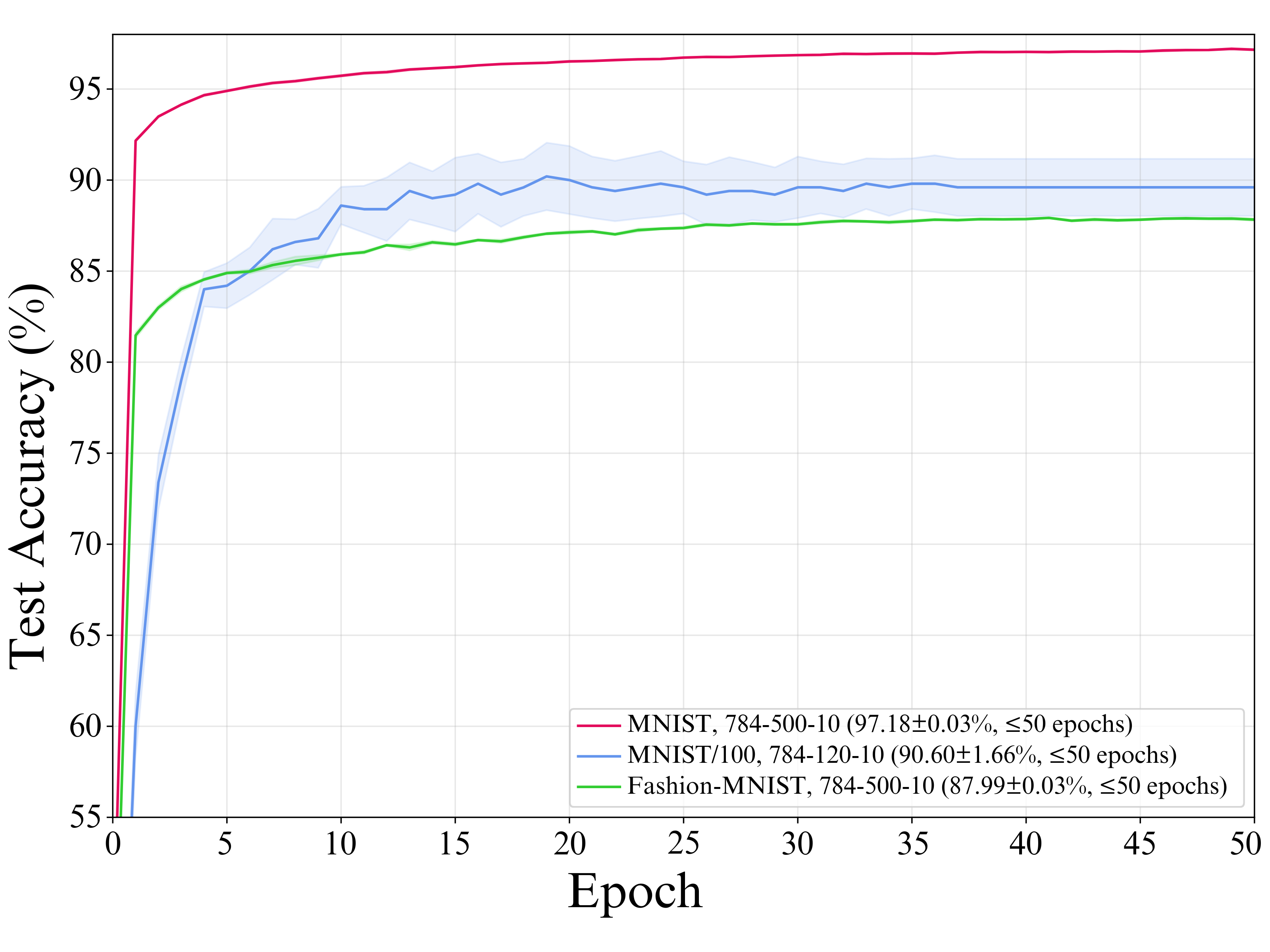}
  
  \vspace{-3mm}
  
  \caption{Test accuracy comparison between OIM architectures on MNIST and Fashion-MNIST up to 50 epochs. The 784-500-10 network achieves \bestMNISTFullAccuracy{} accuracy on full MNIST (red) and \bestFashionMNISTFullAccuracy{} on full Fashion-MNIST (green), while the 784-120-10 network reaches \noiselessBestMNISTHundredAccuracy{} accuracy on MNIST/100 (blue). Shaded regions represent ±1 standard deviation across 5 seeds (often barely visible due to low variance).}

  \label{fig:mnist-fashion-mnist-performance}
  \vspace{-3mm}

\end{figure}

Our OIM networks show strong performance across datasets (Fig.~\ref{fig:mnist-fashion-mnist-performance}), with the 784-120-10 network achieving \noiselessBestMNISTHundredAccuracy{} on MNIST/100 (\noisyBestMNISTHundredAccuracy{} with noise and optimized $\beta$) and the 784-500-10 architecture reaching \bestMNISTFullAccuracy{} accuracy on full MNIST and \bestFashionMNISTFullAccuracy{} on full Fashion-MNIST after 50 epochs.

These results surpass previous hardware implementations, including D-Wave quantum annealers~\cite{laydevantTrainingIsingMachine2024} (88.8±1.5\% on MNIST/100 after 50 epochs), while maintaining competitive accuracy with conventional EP implementations~\cite{ernoultUpdatesEquilibriumProp2019} (97.9±0.1\% on full MNIST after 100 epochs). Performance matches that of BPTT on the same OIM architectures (\bestBPTTMNISTFullAccuracy{} on full MNIST, \bestBPTTMNISTHundredAccuracy{} on MNIST/100), confirming that accuracy is limited by the model architecture rather than the training method. On Fashion-MNIST, our results outperform p-bit Ising machines~\cite{niaziTrainingDeepBoltzmann2024} (85.6\% with 1264 hidden neurons after 120 epochs) and are competitive with other neuromorphic MLP approaches~\cite{zhangTemporalSpikeSequence2020} (89.75±0.03\% with 400-400 hidden neurons after 100 epochs).

\begin{figure}[t]
    \centering
    \includegraphics[width=\columnwidth]{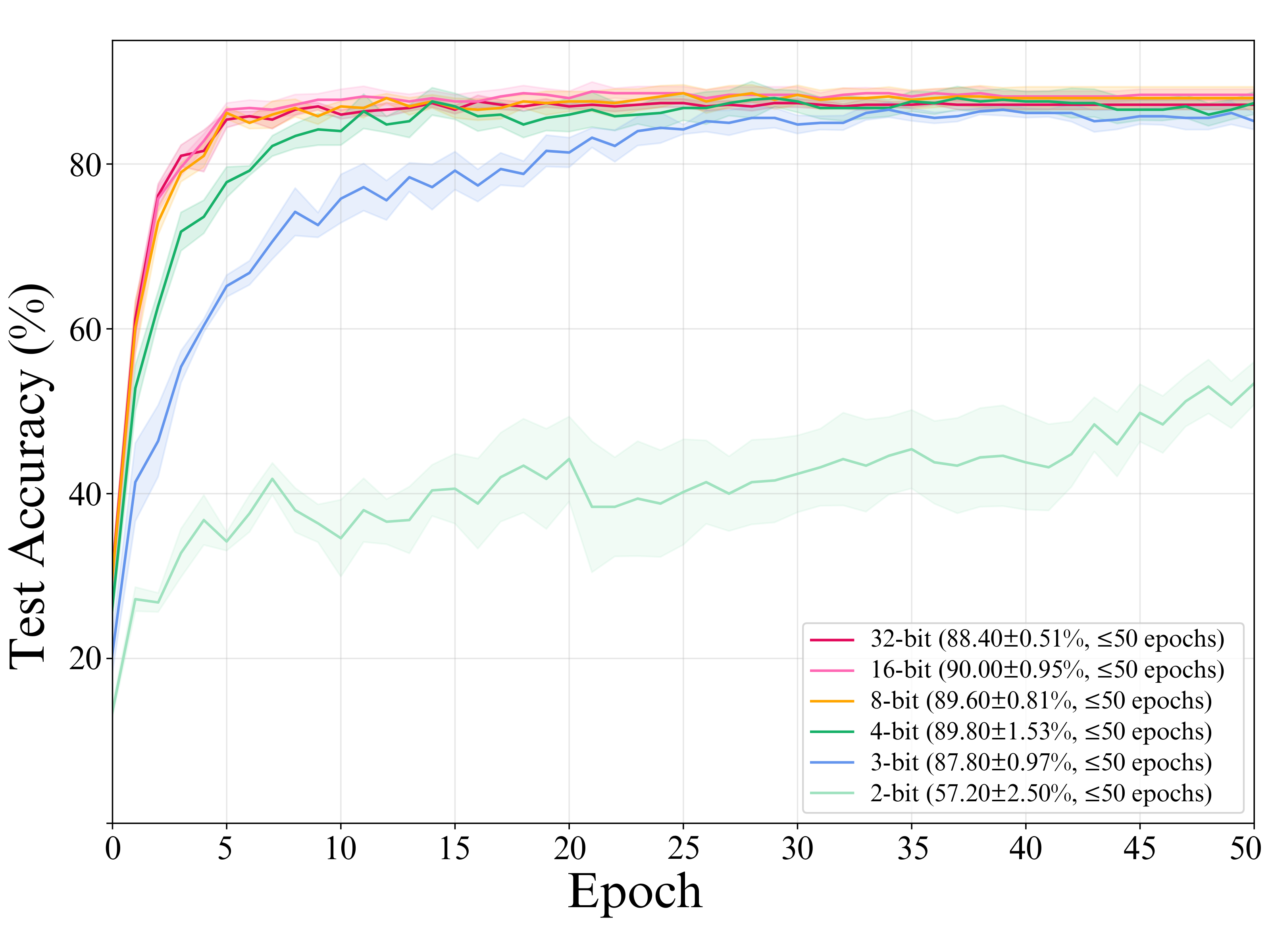}
    \vspace{-3mm}
    \caption{Phase measurement precision effects: Test accuracy vs. training epochs for different phase measurement quantization levels for a 784-120-10 network on MNIST/100. The network maintains full performance down to 4-bit phase measurement precision (\FourBitPhaseQuantisationAccuracy), with significant degradation only at 2-bit precision (\TwoBitPhaseQuantisationAccuracy). This demonstrates that precise phase measurement, while beneficial, is not critical for successful training.}
    \label{fig:phase-measurement-quantization}
    \vspace{-3mm}
\end{figure}

For practical hardware implementation, robustness to physical limitations is essential. Fig.~\ref{fig:phase-measurement-quantization} examines the effects of limited phase measurement precision for EP gradient computation. Here, $n$-bit phase measurement refers to quantizing converged phase values to $2^n$ evenly-spaced values between 0 and $2\pi$: $\phi_*$ before initializing the nudged phases, and $\phi_*^{\pm\beta}$ before applying the EP update equations \eqref{eq:update_bh}--\eqref{eq:update_wxh}. Remarkably, the network maintains full performance down to 4-bit phase measurement (\FourBitPhaseQuantisationAccuracy{} on MNIST/100), with significant degradation only at 2-bit precision.

\begin{figure}[t]
    
    \centering
    
    \includegraphics[width=\columnwidth]{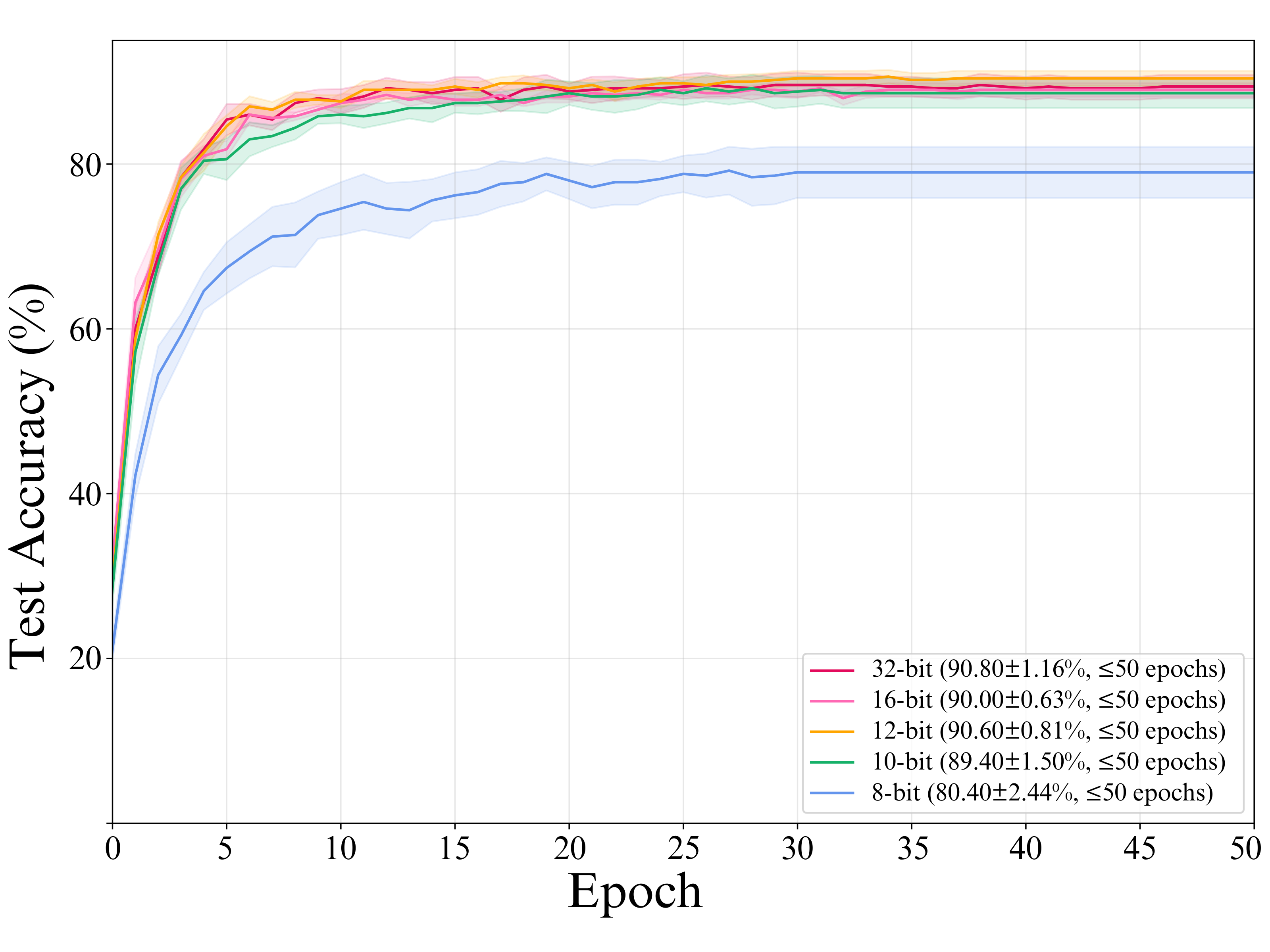}

    \vspace{-3mm}
    
    \caption{Parameter quantization effects: Test accuracy vs. training epochs at different parameter precision levels for a 784-120-10 network on MNIST/100. The network maintains performance with 10-bit precision (\TenBitParameterQuantisationMNISTHundredAccuracy), showing significant degradation only at 8-bit precision (\EightBitParameterQuantisationMNISTHundredAccuracy). This supports hardware implementation with reasonable quantization constraints.}
    
    \label{fig:quantization-effects}
    \vspace{-3mm}

\end{figure}

Similarly, Fig.~\ref{fig:quantization-effects} shows strong performance with 10-bit parameter precision (\TenBitParameterQuantisationMNISTHundredAccuracy{} on MNIST/100), with significant degradation only at 8-bit precision. Here, $n$-bit precision refers to the rounding of coupling, bias, and synchronization parameters to $2^n$ evenly-spaced values within their typical training ranges (determined from full-precision runs) before each dynamics phase. All physical dynamics evolve with these quantized parameters, and gradient updates are computed from the resulting converged phase values, then quantized before application.

 \begin{figure}[t]
    
    \centering
    
    \includegraphics[width=\columnwidth]{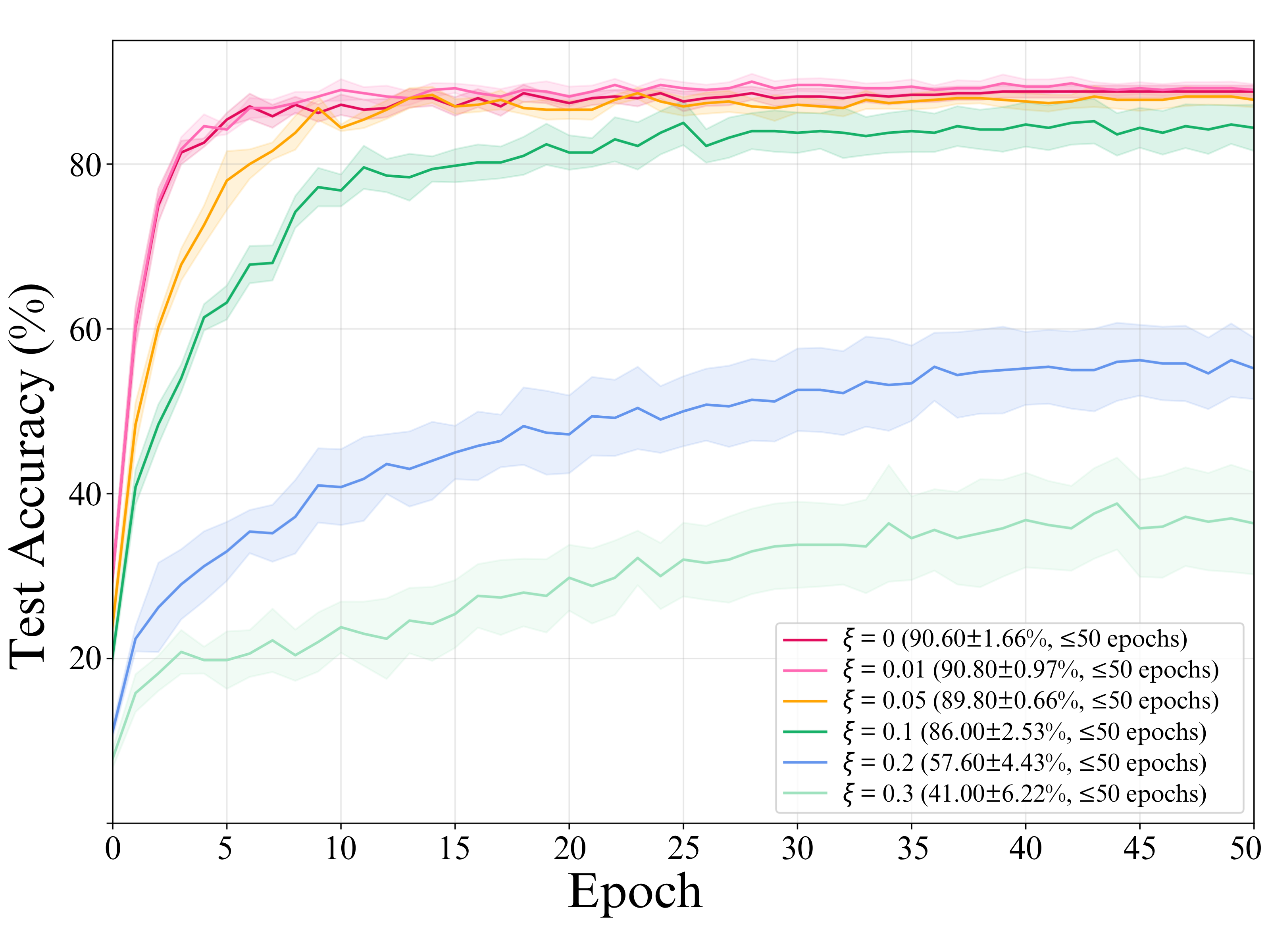}
    \includegraphics[width=\columnwidth]{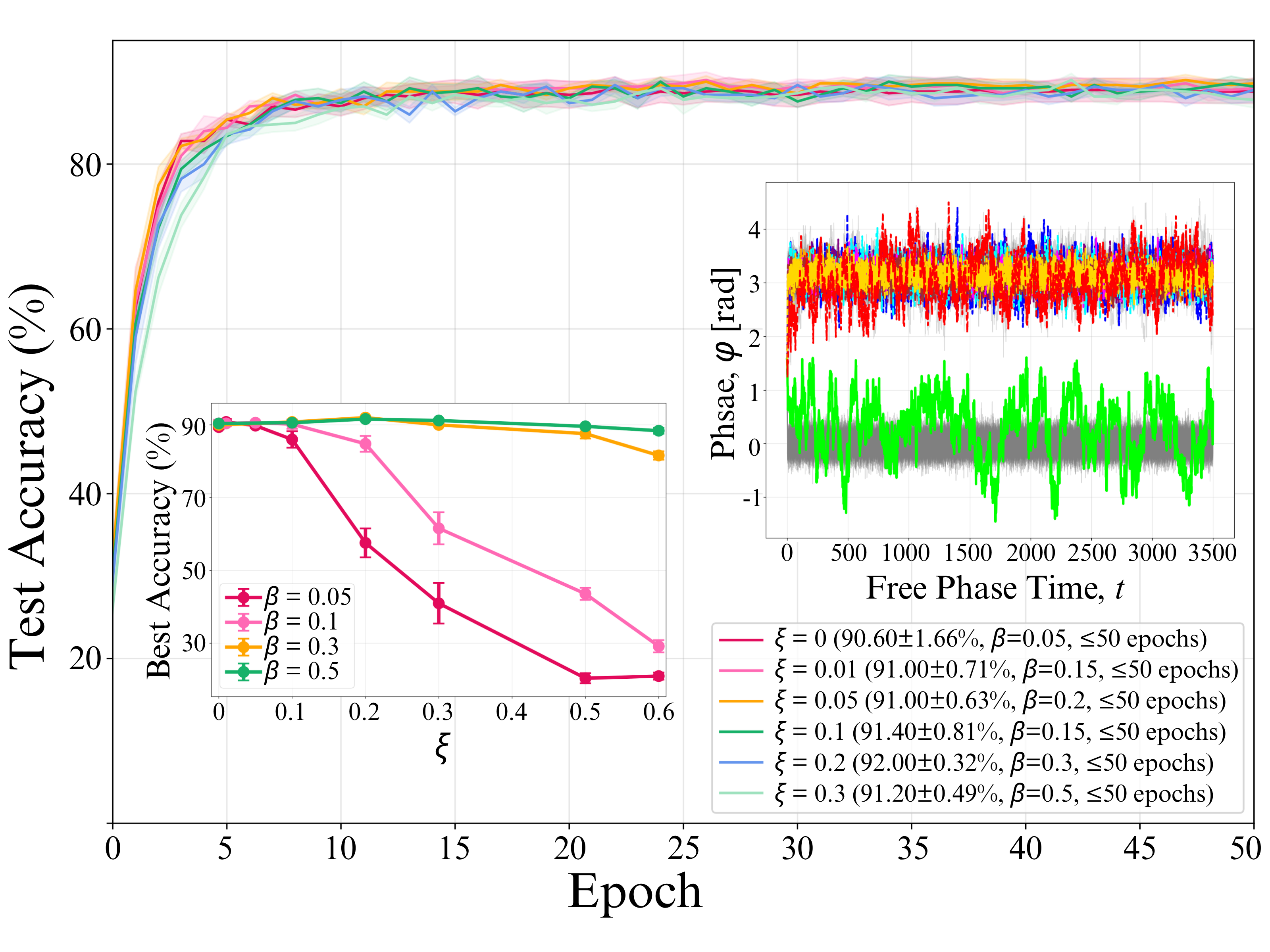}

    \vspace{-3mm}

    \caption{Phase noise effects on training with fixed vs. optimized parameters for a 784-120-10 network on MNIST/100. Top panel: Performance degradation with fixed hyperparameters ($\beta=0.05$) optimized for noise-free conditions, showing decline from \bestDeclineNoiseAccuracy{} to \worstDeclineNoiseAccuracy{} as noise increases. Bottom panel: Performance with noise-optimized nudging factors, achieving peak accuracy of \noisyBestMNISTHundredAccuracy{} at $\xi=0.2$ with $\beta=0.3$, demonstrating that noise can potentially serve as beneficial regularization when paired with appropriately increased nudging factors. Left inset: Performance degradation relationship showing robust performance when $\xi \lesssim 2\beta$. Right Inset: Evolution of oscillator phases over time during a free phase under $\xi=0.3$ noise conditions, with output oscillators (in color) and hidden oscillators (gray).}

    \label{fig:noise-tolerance}
    \vspace{-3mm}

\end{figure}

Fig.~\ref{fig:noise-tolerance} demonstrates noise tolerance through proper parameter tuning. We model phase noise by adding Gaussian noise $\xi \mathcal{N}(0,1)$ to the phase dynamics equation \eqref{eq:oim-dynamics} at each timestep. The top panel shows performance degradation with fixed hyperparameters ($\beta=0.05$) optimized for noise-free conditions, with accuracy declining from \bestDeclineNoiseAccuracy{} to \worstDeclineNoiseAccuracy{} on MNIST/100 as noise increases. However, the bottom panel reveals that when $\beta$ is optimized for each noise level, performance remains robust with no significant degradation and may even improve slightly. The network achieves \noisyBestMNISTHundredAccuracy{} at $\xi=0.2$ with $\beta=0.3$, maintaining performance even at $\xi=0.3$ (unlikely in well-designed implementations, as shown in the right inset). We find that robust performance is maintained when $\beta \gtrsim \xi/2$ (left inset), though theoretical EP guarantees strictly apply in the small $\beta$ limit.

These results confirm viability for physical OIM implementations with realistic hardware constraints: CMOS-based OIMs can achieve 10-bit parameter precision~\cite{royExperimentsOscillatorBased2025}, 4-bit phase detection~\cite{graberIntegratedCoupledOscillator2024}, and moderate phase noise levels that can actually enhance performance when training parameters are optimized.

\section{Discussion}

Our results demonstrate that OIMs can successfully implement Equilibrium Propagation for machine learning tasks with performance comparable to conventional implementations while offering potential benefits in speed, energy efficiency and scalability. The robust performance under realistic hardware constraints—maintaining accuracy with 10-bit parameter precision and 4-bit phase measurement, while potentially benefiting from moderate phase noise—indicates that this approach is viable for physical implementations using current CMOS technology.

Physical OIM implementations offer dramatic speed improvements over software simulations. To estimate this speedup potential: with state-of-the-art GHz-frequency oscillator systems achieving microsecond~\cite{main-OIMs-paper} timescales for parameter setting, convergence, and phase measurement, MNIST training (50 epochs × 60,000 examples × 3 EP phases) could take just seconds to minutes. This compares to $\sim$40 hours for EP and $\sim$60 hours for BPTT using our moderately optimized codebase with the simulation parameters from Table~\ref{tab:hyperparameters}, representing a potential 5-6 order-of-magnitude acceleration. Even with reduced integration steps and lower-frequency oscillators, this would still represent a speedup of several orders of magnitude. This speed advantage could enable practical deployment of energy-based learning methods for computationally intensive tasks that were previously prohibitive due to training time constraints.

The tolerance to moderate noise levels suggests that OIM implementations need not pursue perfect noise suppression. With appropriate $\beta$ tuning ($\beta \gtrsim \xi/2$), robust performance is maintained even at substantial noise levels like $\xi=0.3$, potentially reducing hardware design constraints and implementation costs.

Critically, implementing EP on OIMs requires no major hardware modifications to existing designs. Current OIM implementations already incorporate all essential components—coupling networks, bias fields, synchronization fields, and frequency-calibrated oscillators. This compatibility enables direct repurposing of devices originally designed for combinatorial optimization into neuromorphic processors through software-level parameter configuration and training protocols alone.

\bibliographystyle{IEEEtran}  
\bibliography{references}  

\begin{thebibliography}{10}
\providecommand{\url}[1]{#1}
\csname url@samestyle\endcsname
\providecommand{\newblock}{\relax}
\providecommand{\bibinfo}[2]{#2}
\providecommand{\BIBentrySTDinterwordspacing}{\spaceskip=0pt\relax}
\providecommand{\BIBentryALTinterwordstretchfactor}{4}
\providecommand{\BIBentryALTinterwordspacing}{\spaceskip=\fontdimen2\font plus
\BIBentryALTinterwordstretchfactor\fontdimen3\font minus \fontdimen4\font\relax}
\providecommand{\BIBforeignlanguage}[2]{{%
\expandafter\ifx\csname l@#1\endcsname\relax
\typeout{** WARNING: IEEEtran.bst: No hyphenation pattern has been}%
\typeout{** loaded for the language `#1'. Using the pattern for}%
\typeout{** the default language instead.}%
\else
\language=\csname l@#1\endcsname
\fi
#2}}
\providecommand{\BIBdecl}{\relax}
\BIBdecl

\bibitem{basharNoteAnalyzingStability2023}
M.~K. Bashar, Z.~Lin, and N.~Shukla, ``\BIBforeignlanguage{en}{A note on analyzing the stability of oscillator {Ising} machines},'' \emph{\BIBforeignlanguage{en}{Electronics Letters}}, vol.~59, no.~24, p. e13054, 2023, \_eprint: https://onlinelibrary.wiley.com/doi/pdf/10.1049/ell2.13054.

\bibitem{laydevantTrainingIsingMachine2024}
\BIBentryALTinterwordspacing
J.~Laydevant, D.~Marković, and J.~Grollier, ``\BIBforeignlanguage{en}{Training an {Ising} machine with equilibrium propagation},'' \emph{\BIBforeignlanguage{en}{Nature Communications}}, vol.~15, no.~1, p. 3671, Apr. 2024, publisher: Nature Publishing Group. [Online]. Available: \url{https://www.nature.com/articles/s41467-024-46879-4}
\BIBentrySTDinterwordspacing

\bibitem{rageauTrainingSynchronizingOscillator2025}
\BIBentryALTinterwordspacing
T.~Rageau and J.~Grollier, ``\BIBforeignlanguage{en}{Training and synchronizing oscillator networks with {Equilibrium} {Propagation}},'' \emph{\BIBforeignlanguage{en}{Neuromorphic Computing and Engineering}}, vol.~5, no.~3, p. 034008, Jul. 2025, publisher: IOP Publishing. [Online]. Available: \url{https://dx.doi.org/10.1088/2634-4386/adebaa}
\BIBentrySTDinterwordspacing

\bibitem{garcia-martinEstimationEnergyConsumption2019}
\BIBentryALTinterwordspacing
E.~García-Martín, C.~F. Rodrigues, G.~Riley, and H.~Grahn, ``\BIBforeignlanguage{en}{Estimation of energy consumption in machine learning},'' \emph{\BIBforeignlanguage{en}{Journal of Parallel and Distributed Computing}}, vol. 134, pp. 75--88, Dec. 2019. [Online]. Available: \url{https://linkinghub.elsevier.com/retrieve/pii/S0743731518308773}
\BIBentrySTDinterwordspacing

\bibitem{csabaCoupledOscillatorsComputing2020b}
\BIBentryALTinterwordspacing
G.~Csaba and W.~Porod, ``Coupled oscillators for computing: {A} review and perspective,'' \emph{Applied Physics Reviews}, vol.~7, no.~1, p. 011302, Jan. 2020. [Online]. Available: \url{https://doi.org/10.1063/1.5120412}
\BIBentrySTDinterwordspacing

\bibitem{assranSelfSupervisedLearningImages2023}
\BIBentryALTinterwordspacing
M.~Assran, Q.~Duval, I.~Misra, P.~Bojanowski, P.~Vincent, M.~Rabbat, Y.~LeCun, and N.~Ballas, ``Self-{Supervised} {Learning} from {Images} with a {Joint}-{Embedding} {Predictive} {Architecture},'' Apr. 2023, arXiv:2301.08243 [cs]. [Online]. Available: \url{http://arxiv.org/abs/2301.08243}
\BIBentrySTDinterwordspacing

\bibitem{wangTrainingCoupledPhase2024}
\BIBentryALTinterwordspacing
Q.~Wang, C.~C. Wanjura, and F.~Marquardt, ``Training {Coupled} {Phase} {Oscillators} as a {Neuromorphic} {Platform} using {Equilibrium} {Propagation},'' Feb. 2024, arXiv:2402.08579 [cond-mat, physics:physics]. [Online]. Available: \url{http://arxiv.org/abs/2402.08579}
\BIBentrySTDinterwordspacing

\bibitem{main-OIMs-paper}
T.~Wang, L.~Wu, P.~Nobel, and J.~Roychowdhury, ``\BIBforeignlanguage{en}{Solving combinatorial optimisation problems using oscillator based {Ising} machines},'' \emph{\BIBforeignlanguage{en}{Natural Computing}}, vol.~20, no.~2, pp. 287--306, Jun. 2021.

\bibitem{scellierEquilibriumPropagationBridging2017}
\BIBentryALTinterwordspacing
B.~Scellier and Y.~Bengio, ``Equilibrium {Propagation}: {Bridging} the {Gap} {Between} {Energy}-{Based} {Models} and {Backpropagation},'' Mar. 2017, arXiv:1602.05179. [Online]. Available: \url{http://arxiv.org/abs/1602.05179}
\BIBentrySTDinterwordspacing

\bibitem{laborieuxScalingEquilibriumPropagation2020}
\BIBentryALTinterwordspacing
A.~Laborieux, M.~Ernoult, B.~Scellier, Y.~Bengio, J.~Grollier, and D.~Querlioz, ``Scaling {Equilibrium} {Propagation} to {Deep} {ConvNets} by {Drastically} {Reducing} its {Gradient} {Estimator} {Bias},'' Jun. 2020, arXiv:2006.03824. [Online]. Available: \url{http://arxiv.org/abs/2006.03824}
\BIBentrySTDinterwordspacing

\bibitem{werbosBackpropagationTimeWhat1990}
\BIBentryALTinterwordspacing
P.~Werbos, ``Backpropagation through time: what it does and how to do it,'' \emph{Proceedings of the IEEE}, vol.~78, no.~10, pp. 1550--1560, Oct. 1990. [Online]. Available: \url{https://ieeexplore.ieee.org/document/58337}
\BIBentrySTDinterwordspacing

\bibitem{heDelvingDeepRectifiers2015}
\BIBentryALTinterwordspacing
K.~He, X.~Zhang, S.~Ren, and J.~Sun, ``Delving {Deep} into {Rectifiers}: {Surpassing} {Human}-{Level} {Performance} on {ImageNet} {Classification},'' Feb. 2015, arXiv:1502.01852 [cs]. [Online]. Available: \url{http://arxiv.org/abs/1502.01852}
\BIBentrySTDinterwordspacing

\bibitem{lecunGradientbasedLearningApplied1998}
\BIBentryALTinterwordspacing
Y.~Lecun, L.~Bottou, Y.~Bengio, and P.~Haffner, ``Gradient-based learning applied to document recognition,'' \emph{Proceedings of the IEEE}, vol.~86, no.~11, pp. 2278--2324, Nov. 1998. [Online]. Available: \url{https://ieeexplore.ieee.org/document/726791}
\BIBentrySTDinterwordspacing

\bibitem{xiaoFashionMNISTNovelImage2017}
\BIBentryALTinterwordspacing
H.~Xiao, K.~Rasul, and R.~Vollgraf, ``Fashion-{MNIST}: a {Novel} {Image} {Dataset} for {Benchmarking} {Machine} {Learning} {Algorithms},'' Aug. 2017, aDS Bibcode: 2017arXiv170807747X. [Online]. Available: \url{https://ui.adsabs.harvard.edu/abs/2017arXiv170807747X}
\BIBentrySTDinterwordspacing

\bibitem{ernoultUpdatesEquilibriumProp2019}
\BIBentryALTinterwordspacing
M.~Ernoult, J.~Grollier, D.~Querlioz, Y.~Bengio, and B.~Scellier, ``Updates of {Equilibrium} {Prop} {Match} {Gradients} of {Backprop} {Through} {Time} in an {RNN} with {Static} {Input},'' May 2019, arXiv:1905.13633 [cs]. [Online]. Available: \url{http://arxiv.org/abs/1905.13633}
\BIBentrySTDinterwordspacing

\bibitem{niaziTrainingDeepBoltzmann2024}
\BIBentryALTinterwordspacing
S.~Niazi, N.~A. Aadit, M.~Mohseni, S.~Chowdhury, Y.~Qin, and K.~Y. Camsari, ``Training {Deep} {Boltzmann} {Networks} with {Sparse} {Ising} {Machines},'' \emph{Nature Electronics}, vol.~7, no.~7, pp. 610--619, Jun. 2024, arXiv:2303.10728 [cs]. [Online]. Available: \url{http://arxiv.org/abs/2303.10728}
\BIBentrySTDinterwordspacing

\bibitem{zhangTemporalSpikeSequence2020}
\BIBentryALTinterwordspacing
W.~Zhang and P.~Li, ``Temporal {Spike} {Sequence} {Learning} via {Backpropagation} for {Deep} {Spiking} {Neural} {Networks},'' in \emph{Advances in {Neural} {Information} {Processing} {Systems}}, vol.~33.\hskip 1em plus 0.5em minus 0.4em\relax Curran Associates, Inc., 2020, pp. 12\,022--12\,033. [Online]. Available: \url{https://proceedings.neurips.cc/paper/2020/hash/8bdb5058376143fa358981954e7626b8-Abstract.html}
\BIBentrySTDinterwordspacing

\bibitem{royExperimentsOscillatorBased2025}
\BIBentryALTinterwordspacing
S.~Roy and B.~Ulmann, ``Experiments with an oscillator based {Ising} machine,'' Feb. 2025, arXiv:2502.03167 [cs] version: 1. [Online]. Available: \url{http://arxiv.org/abs/2502.03167}
\BIBentrySTDinterwordspacing

\bibitem{graberIntegratedCoupledOscillator2024}
\BIBentryALTinterwordspacing
M.~Graber and K.~Hofmann, ``\BIBforeignlanguage{en}{An integrated coupled oscillator network to solve optimization problems},'' \emph{\BIBforeignlanguage{en}{Communications Engineering}}, vol.~3, no.~1, p. 116, Aug. 2024, publisher: Nature Publishing Group. [Online]. Available: \url{https://www.nature.com/articles/s44172-024-00261-w}
\BIBentrySTDinterwordspacing

\end{thebibliography}

\end{document}